\documentclass[12pt]{article}
\usepackage{epsfig,amsfonts,amsthm}
\newcommand{\be}{\begin{equation}}
\newcommand{\ee}{\end{equation}}
\newcommand{\bea}{\begin{eqnarray}}
\newcommand{\eea}{\end{eqnarray}}
\newcommand{\dst}{\displaystyle}
\newcommand{\fr}[2]{\frac{{\dst #1}}{{\dst #2}}}

\newcommand{\fd}{\phi^\dagger}

\newcommand{\stolb}[3]{ \left( \begin{array}{c}#1 \\ #2 \\ #3\end{array}\right) }

\def\lsim{\mathrel{\rlap{\lower4pt\hbox{\hskip1pt$\sim$}}
    \raise1pt\hbox{$<$}}}         
\def\gsim{\mathrel{\rlap{\lower4pt\hbox{\hskip1pt$\sim$}}
    \raise1pt\hbox{$>$}}}         

\title{Can 2HDM support fermion-stabilized bubbles of false vacuum?}

\author{I.P. Ivanov\thanks{E-mail: Igor.Ivanov@ulg.ac.be}\\
  {\small Interactions Fondamentales en Physique et en Astrophysique, Universit\'{e} de Li\`{e}ge,} \\
  {\small All\'{e}e du 6 Ao\^{u}t, 17, b\^{a}timent B5a, B-4000 Li\`{e}ge, Belgium}\\
  {\small and}\\
  {\small Sobolev Institute of Mathematics, ac. Koptyug av. 4, 630090, Novosibirsk, Russia}}
\date{}

\begin{document}
\maketitle

\begin{abstract}
The Higgs potential of the two-Higgs-doublet model
can have several minima with different properties.
We discuss a possibility that a heavy fermion, if trapped in a microscopic
false vacuum bubble, might become light enough to prevent the bubble from the collapse.
\end{abstract}

The Standard Model relies on the Higgs mechanism of the
electroweak symmetry breaking, whose details are not yet
known. Its simplest realization is based
on a single weak isodoublet of scalar fields, while extended
versions deal with more elaborate scalar sectors, see reviews
\cite{Hunter,workshop}.
The two-Higgs-doublet model \cite{2HDM}, where one
introduces two Higgs doublets $\phi_1$ and $\phi_2$, is one of the
most economic extensions of the Higgs sector beyond the Standard
Model. The minimal supersymmetric extension of
the Standard Model (MSSM) uses precisely a specific version of the
2HDM to break the electroweak symmetry.

The Higgs potential of the most general 2HDM $V_H = V_2 + V_4$ is
conventionally parametrized as
\bea V_2&=&-{1\over 2}\left[m_{11}^2(\phi_1^\dagger\phi_1) +
m_{22}^2(\phi_2^\dagger\phi_2)
+ m_{12}^2 (\phi_1^\dagger\phi_2) + m_{12}^{2\ *} (\phi_2^\dagger\phi_1)\right]\,;\nonumber\\
V_4&=&\fr{\lambda_1}{2}(\phi_1^\dagger\phi_1)^2
+\fr{\lambda_2}{2}(\phi_2^\dagger\phi_2)^2
+\lambda_3(\phi_1^\dagger\phi_1) (\phi_2^\dagger\phi_2)
+\lambda_4(\phi_1^\dagger\phi_2) (\phi_2^\dagger\phi_1) \label{potential}\\
&+&\fr{1}{2}\left[\lambda_5(\phi_1^\dagger\phi_2)^2+
\lambda_5^*(\phi_2^\dagger\phi_1)^2\right]
+\left\{\left[\lambda_6(\phi_1^\dagger\phi_1)+\lambda_7
(\phi_2^\dagger\phi_2)\right](\phi_1^\dagger\phi_2) +{\rm h.c.}\right\}\,.\nonumber
\eea
It contains 14 free parameters: real
$m_{11}^2, m_{22}^2, \lambda_1, \lambda_2, \lambda_3, \lambda_4$
and complex $m_{12}^2, \lambda_5, \lambda_6, \lambda_7$.

The large number of free parameters makes the analysis of the most general
2HDM and its phenomenological consequences rather complicated.
Even the first step, straightforward minimization of the potential
is prohibitively difficult for the most general 2HDM.
This is why the most activity so far has been focused
on some particular version of 2HDM with a strongly reduced number of free parameters.

On the other hand, the most general 2HDM has a remarkably rich spectrum of phases,
which is lost in many specific version of 2HDM. The number of extrema, their nature,
and various phase transitions in the most general 2HDM
can be studied without explicit algebraic minimization of the potential, within
the geometric approach developed in \cite{mink}.
This approach helps easily see various possibilities offered by the 2HDM.

In this Letter we discuss a particular possibility that might be realized in 2HDM:
a microscopic bubble of the false vacuum stabilized by a heavy fermion trapped in it.

The idea goes as follows. 2HDM can have several minima with different properties \cite{mink,different}.
One assumes that our world is in the global minimum (true vacuum).
If a false vacuum bubbles is produced,
it quickly collapses due to the large surface tension \cite{bubbles}.
However, one can construct a 2HDM, in which a fermion that is heavy in our world
can become much lighter inside the false vacuum bubble.
This mass difference might be sufficient to compensate the extra energy associated
with the bubble surface tension, thus stabilizing the bubble against the collapse.

Here, we present quasiclassical estimates indicating that such a possibility should not be neglected.
It is of course difficult to say how reliable these quasiclassical estimates of
the properties of {\em microscopic} bubbles are.
A more accurate analysis is needed to establish whether it can be indeed realized in 2HDM
without violation of the experimental bounds.
If 2HDM indeed supports such a possibility, it will have remarkable phenomenological consequences.\\

Let us first describe the construction of a 2HDM that contains two nearly degenerate
minima with almost equal $v_1^2 + v_2^2$ but very different fermion masses.
We start with the compact representation of the Higgs potential of 2HDM introduced in
\cite{mink,nishi}:
\be
V = - M_\mu r^\mu + {1\over 2}\Lambda_{\mu\nu} r^\mu r^\nu\,.\label{Vmunu}
\ee
Here the Higgs fields appear as components of a four-vector $r^\mu = (r^0,\,r^i)$ with
\be
r_0 = (\Phi^\dagger \Phi) = (\fd_1 \phi_1) + (\fd_2 \phi_2)\,,\quad
r_i = (\Phi^\dagger \sigma_i \Phi) =
\stolb{(\fd_2 \phi_1) + (\fd_1 \phi_2)}{-i[(\fd_1 \phi_2) - (\fd_2 \phi_1)]}{(\fd_1 \phi_1) - (\fd_2 \phi_2)}\,,
\label{ri}
\ee
and the neutral vacuum we consider corresponds to vacuum points $r^\mu$ lying on the
forward lightcone, $r^\mu r_\mu = 0$.
Thus, the orbit space of the 2HDM Higgs potential is equipped with the Minkowski space structure.
The group of special linear transformations of doublets $\phi_1$ and $\phi_2$ induces
the Lorentz group $SO(1,3)$ of transformations of $r^\mu$.

The ten quartic couplings $\lambda_i$ of the most general 2HDM define the
real symmetric tensor $\Lambda_{\mu\nu}$, while the four free parameters
of the quadratic term form the real four-vector $M_\mu$.
The explicit expressions for $\Lambda_{\mu\nu}$ and $M_\mu$ can be found in \cite{mink,nishi}.
Here we mention only the fact that $\Lambda_{\mu\nu}$ can be always diagonalized by an $SO(1,3)$
transformation. To simplify the discussion we assume that $\Lambda_{\mu\nu}$ is already diagonal.

One of the features of 2HDM is the possibility to spontaneously break discrete symmetries.
The most studied case is the spontaneous $CP$-violation in the Higgs sector, \cite{CP}.
It corresponds to the situation when $M_2 = 0$ but $r^\mu$ has a non-zero component $r_2$, which according
to (\ref{ri}) corresponds to a nonzero relative phase between $v_1$ and $v_2$.

\begin{figure}[!htb]
   \centering
\includegraphics[width=10cm]{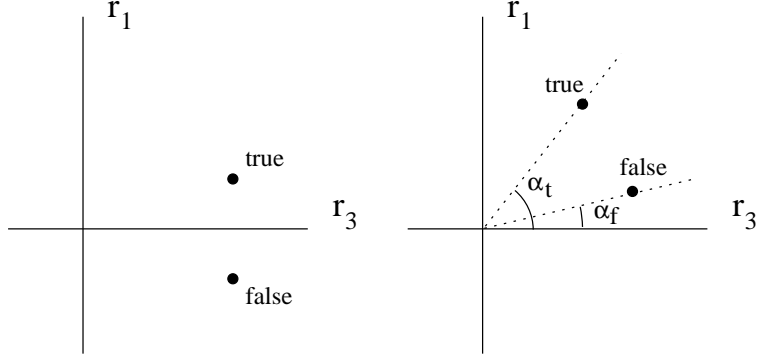}
\caption{Location of the minima in the orbit space of a specific 2HDM. One starts from
a nearly degenerate situation with, right, and then rotates the construction
to obtain model discussed in the text, left.}
   \label{fig1}
\end{figure}

Analogously, one can consider a situation when $M_1=0$, but $r_1 \not = 0$ (see Fig.~\ref{fig1}, left).
It corresponds to a situation when the Higgs potential respects $Z_2$ symmetry (i.e.
symmetry under sign change of one of the doublets), but it is spontaneously violated
in each of the two degenerate minima (i.e. minima correspond to non-zero v.e.v.'s
of both doublets).
The requirements for such a situation to take place is that $M_\mu$ must lie inside
the corresponding caustic cone in the Minkowski space (see details in \cite{mink}).
For simplicity, we also take $M_2=0$ and $r_2=0$.
Note that $v_0^2 = v_1^2+v_2^2 = (246 \mathrm{GeV})^2$ is the same for both minima.
In addition, one can always choose
parameters $\lambda_i$ such that there is no other minimum of the potential.
All these criteria are realized, for instance, in the 2HDM with the following parameters
\bea
&&\lambda_1=\lambda_2>0\,,\quad \lambda_5< 0\,,\quad |\lambda_5|>\lambda_4\,,\quad\lambda_3 = \lambda_6=\lambda_7=0\,,\nonumber\\
&& m_{11}^2,\, m_{22}^2 > 0\,,\quad m_{11}^2\not = m_{22}^2\,,\quad m_{12}^2 = 0\,.
\label{example}
\eea

One can remove degeneracy by slightly shifting $M^\mu$ away from
the $\Lambda_{\mu\nu}$ eigenaxis (by taking small but nonzero Re\,$m_{12}^2$),
which makes one of the minima (the true vacuum) slightly deeper
than the other. This difference can be made small enough so that it does not significantly affect
the values of $v_i$ and the energetic considerations below.

Now we rotate this construction to obtain the one shown in Fig.~\ref{fig1}, right.
This can be done by an appropriate
change of parameters of the Higgs potential, for example, by introducing some real $\lambda_6 = -\lambda_7$
together with matching Re\,$m_{12}^2$. The two nearly degenerate minima now correspond to
\bea
&&\mbox{true vacuum:}\quad v_1 = v_0\cos\beta_t,\quad v_2=v_0\sin\beta_t\equiv v_t\,;\nonumber\\
&&\mbox{false vacuum:}\quad v_1 = v_0\cos\beta_f,\quad v_2=v_0\sin\beta_f\equiv v_f\,.
\eea
Angles $\beta_t$ and $\beta_f$ can be chosen at will. The choice we make is
$\beta_t$ being noticably larger than $\beta_f$.
Angles $\alpha_i$ shown in Fig.~\ref{fig1} are twice $\beta_i$.

Let us now suppose that the heavy fermion gets its mass via coupling only
to the second doublet. In the true and false vacuum we have, respectively,
\be
m_t = g {v_t\over \sqrt{2}}\,,\quad m_f = g {v_f\over \sqrt{2}}\,.\label{mtmf}
\ee

Let us now focus on the Higgs potential along the least energy path in the Higgs space that
connects the two minima. If the minima are close to each other, we can approximate the path by a straight line.
Let $h$ be the neutral scalar field along this direction, whose zero is taken just between the minima.
Then the Higgs lagrangian density is approximately
\be
L = {1\over 2}(\partial^\mu h)(\partial_\mu h) - {\lambda \over 2}\left[{1\over 2}h^2
-{1\over 2}\left({v_t-v_f \over 2}\right)^2 \right]^2 + L_0\,,
\ee
where $L_0$ includes all other terms and does not significantly change between the two minima.
Here $\lambda$ is the coefficient of the quartic potential along the chosen direction.
The Higgs boson mass along this direction in each of the two minima is
\be
m_h = \sqrt{\lambda}\,{v_t-v_f \over 2}\,.\label{mh}
\ee
We need to estimate the energy associated with the corresponding barrier.

Consider now a spherical bubble of the false vacuum of radius $\xi$. We take this radius large enough to
allow the fermion (with its false vacuum mass!) to fit into the bubble: $\xi = 1/m_f$.
A typical radial field profile of the bubble is the following:
false vacuum at $r\lsim \xi$, barrier around $r \sim \xi$ and the true vacuum at $r\gsim \xi$.
For estimate, we assume that the barrier wall spans from
$r=\xi/2$ to $r=3\xi/2$ and that the field changes by about $(v_t-v_f)/2$ over this distance.
Then, the energy associated with the static bubble of such a field profile is approximately
\bea
\Delta E_H &\sim & {4 \pi\over 3} \left[\left({3 \over 2}\right)^3 - \left({1 \over 2}\right)^3\right]\xi^3
\cdot {1\over 2}\left[\left({v_t-v_f\over 2}\right)^2{1\over \xi^2}
+{\lambda \over 4}\left({v_t-v_f\over 2}\right)^4\right]\nonumber\\
&= & {13\pi \over 6}\xi \left({v_t-v_f\over 2}\right)^2
\left[1 +{\lambda \over 4}\left({v_t-v_f\over 2}\right)^2\xi^2\right]\,.
\eea
The stability condition is that the fermion mass difference between the true and false vacua
be larger than this barrier energy
\be
\Delta E_f = g {v_t-v_f\over \sqrt{2}} > \Delta E_H \,.
\ee
Substituting $\xi = \sqrt{2}/(g v_f)$, we get
\be
g^2 > {13\pi \over 12} x \left(1+{\lambda \over 8g^2}x^2\right)\,,
\quad\mbox{where}\quad x \equiv {v_t-v_f \over v_f}\,.
\label{limit}
\ee
We now consider two options for the heavy fermion: the top-quark and a new very heavy fermion.
\\

{\bf Top quark.}
The natural candidate for the heavy fermion would be the top-quark.
Satisfying this inequality without generating too light Higgs boson implies not only large $g$,
but also large $\lambda$, as can be seen from (\ref{mtmf}) together with (\ref{mh}).
The exact experimental limit on the lightest Higgs mass is not known in the
most general 2HDM, so we take the PDG bound on the lightest supersymmetric Higgs, $m_H > 89.8$ GeV, \cite{bound}.
Then the above inequality can be satisfied, for example, with
$\lambda=12$, $g=2.2$, and $x=1$ so that $v_t = 2 v_f = 0.447 v_0$ and $m_H = 95$ GeV.
This implies, however, that the bubble itself is rather light, $\sim 80$ GeV with this particular
choice of parameters. It is even lighter than the Higgs boson itself,
which seems to be a very unlikely possibility.

According to \cite{unitary}, where the tree-level unitarity constraints
were presented for the most general 2HDM, values of $\lambda$ larger than $\sim 10$
lead to intense coupling regime in the Higgs sector,
where the conclusions based on the tree-level analysis are hardly reliable.
However, the above estimate is very sensitive to numerical factors.
It remains to be seen if a more accurate calculation of $\Delta E_H$
can make $\lambda$ smaller.

Let us briefly discuss the phenomenological properties of such a bubble
if it can be indeed realized in 2HDM. This bubble should
manifest itself as a particle with a mass below the known $t$-quark mass.
Its lifetime will depend on the stability of the fermion
in the false vacuum (the bubble cannot collapse before the fermion decays).
Since its mass in the false vacuum is several times smaller than in the true vacuum,
its width should be below 1 GeV scale. Not only the smaller mass reduces the phase space, but it also makes its decay
to $bW^+$ subthreshold (the $W$ mass does not change significantly in the false vacuum considered,
since $v_0$ is kept the same). Its production at colliders should be a rare
process since an energetic collision must produce simultaneously a top-quark
{\em and} a specific Higgs field configuration around it.\\

{\bf New very heavy fermions.}
If new very heavy fermions exist and if their masses are also generated by the Higgs
mechanism, they will readily provide a large enough Yukawa constant $g$ to satisfy (\ref{limit}).
A very heavy fermion also allows for a sufficiently heavy bubble.
However, it will also lead to large loop corrections to the Higgs potential,
so it remains to be seen whether how strongly the above estimates get modified.
\\

{\bf Acknowledgements}. I am thankful to Ilya Ginzburg for stimulating discussions of various aspects
of 2HDM. The work was supported by FNRS and partly by grants RFBR 05-02-16211 and NSh-5362.2006.2.

\end{document}